\documentclass[useAMS,usenatbib]{mn2e}

\usepackage{graphicx,natbib,psfig}
\usepackage{amsmath}
\usepackage{amssymb}

\def\bc{\begin{center}}
\def\ec{\end{center}}

\title[Searches for UCDs in Galaxy Groups]{Searches for Ultra-Compact Dwarf Galaxies \\ in Galaxy Groups}
\author[Evstigneeva et al.]
{E.~A.~Evstigneeva$^{1,6}$, M.~J.~Drinkwater$^{1}$, R.~Jurek$^{1}$, P.~Firth$^{1}$, J.~B.~Jones$^{2}$, \and M.~D.~Gregg$^{3,4}$, S.~Phillipps$^{5}$\\ \\
$^1$Department of Physics, University of Queensland, QLD 4072, Australia\\
$^2$Astronomy Unit, School of Mathematical Sciences, Queen Mary University of London, Mile End Road, London, E1 4NS, UK\\
$^3$Department of Physics, University of California, Davis, CA 95616, USA\\
$^4$Institute for Geophysics and Planetary Physics, Lawrence Livermore National Laboratory, L-413, Livermore, CA 94550, USA\\
$^5$Astrophysics Group, Department of Physics, University of Bristol, Tyndall Avenue, Bristol, BS8 1TL, UK\\
$^6$E-mail: katya@physics.uq.edu.au}

\pagerange{000-000} \pubyear{2007}

\begin{document}

\maketitle

\begin{abstract}
We present the results of a search for ultra-compact dwarf galaxies (UCDs)
in six different galaxy groups: Dorado, NGC1400, NGC0681, NGC4038, NGC4697 and NGC5084. 
We searched in the apparent magnitude range $17.5 \leq b_j \leq 20.5$ 
(except NGC5084: $19.2 \leq b_j \leq 21.0$). We found 1 definite plus
2 possible UCD candidates in the Dorado group and 2 possible UCD candidates in
the NGC1400 group. No UCDs were found in the other groups. 
We compared these results with predicted luminosities of UCDs in the
groups according to the hypothesis that UCDs are globular clusters formed in galaxies. 
The theoretical predictions broadly agree with the observational results, but deeper surveys are needed 
to fully test the predictions.
\end{abstract}

\begin{keywords}
astronomical data bases: surveys -- galaxies: distances and redshifts -- galaxies: dwarf -- 
galaxies: star clusters 
\end{keywords}

\section{Introduction}

Ultra-compact dwarf galaxies (UCDs) have recently been proposed as a new type of 
stellar system. They were discovered independently by Hilker et al. (1999) (2 objects) and 
Drinkwater et al. (2000a) (5 objects including 2 Hilker's objects) in spectroscopic surveys 
of the centre of the Fornax Cluster.
The discovered objects have absolute magnitudes in the range $-13.5 \lesssim M_B \lesssim -11.5$,  
properties intermediate between globular clusters (GCs) and dwarf galaxies, 
and are mostly unresolved in ground-based imaging (half-light radii $<100$ pc). 
Searches for similar objects in the Virgo Cluster have revealed a population of 9 confirmed UCDs 
in the centre of the cluster (Jones et al. 2006).
Further spectroscopic surveying of the Fornax Cluster by Drinkwater et al. (2004), 1.5 mag deeper than the
original observations (Drinkwater et al. 2000a), has found more than 50 new UCDs in a 0$\degr$.9 radius field 
centered on the first ranked galaxy NGC1399. Mieske, Hilker \& Infante (2004a), in their spectroscopic study of the Fornax Cluster,  
identified 54 ultra-compact objects with magnitudes down to $V < 21.0$ ($M_B \approx -9.5$ mag) 
within $\sim 20\arcmin$ of NGC1399. 
UCD candidates were also found from HST imaging of the more distant cluster Abell~1689 (Mieske et al. 2004b).

The main formation scenarios for UCDs are as follows. i) They are very massive
(intra-cluster) globular clusters (Hilker et al. 1999; Drinkwater et al. 2000a; 
Phillipps et al. 2001; Mieske, Hilker \& Infante 2002; Evstigneeva et al. 2007). ii) They are 
the remnant nuclei of stripped (threshed) early-type dwarf galaxies (Bekki, Couch \& Drinkwater 2001; Bekki et al. 2003). 
iii) They are evolved products of YMGCs (young massive GCs) -- massive super starclusters 
formed in galaxy interactions (Fellhauer \& Kroupa 2002; Maraston et al. 2004).

To date, confirmed UCDs have only been found in the centers of rich
galaxy clusters. The aim of this work is to determine if UCDs exist in
less dense environments such as galaxy groups and, if they do, to
compare their properties to Fornax and Virgo UCDs.  Previous studies
of UCDs in galaxy groups are limited to the photometric search for
UCDs in the NGC1023 group by Mieske, West \& Mendes de Oliveira (2006). 21 possible UCD
candidates were found. Mieske et al. showed that the mass spectrum of
the UCD candidates in NGC1023 is restricted to $\sim 1/4$ of the
maximum Fornax and Virgo UCD mass. However, spectroscopy is required
to confirm UCD candidates in the NGC1023 group.

In this paper we present a spectroscopic search for UCDs in a range
of galaxy group environments. Identifying UCDs and defining their
properties in different environments can help us to put constraints on
UCD formation mechanisms.

\begin{table*}
\caption[1]{Galaxy groups. The numbers for the Fornax and Virgo Clusters are given for a comparison.} 
\bc
\begin{tabular}{lcclcc}
\hline
\\
 & Distance & \multispan3{\hfil {\it Most luminous galaxies in 2dF field}\hfil} & Number of spectroscopically \\ 
Group & modulus & Name & Type & $M_B$ & confirmed members\\ 
 & (mag) & & & (mag) & within 2dF field \\
 & & & & &  \\
(1) & (2) & (3) & (4) & (5) & (6)  \\
\hline
\\
Dorado & 30.92 & NGC1553 & S0 & -20.71 & 11 \\
 & & NGC1549 & E & -20.31 &  \\ \\
NGC~1400/1407 & 31.46 & NGC1407 & E & -21.07 & 34 \\
 & & NGC1400 & E-S0 & -19.80 &  \\ \\
NGC~0681 & 31.51 & NGC0681 & SABa & -19.12 & 10 \\ 
 & & NGC0701 & SBc & -19.37 &  \\ \\ 
NGC~4038 & 32.26 & NGC4038 & SBm & -21.99 & 10 \\
 & & NGC4039 & SBm & -21.81 & \\ 
 & & NGC4027 & SBd & -21.13 & \\ \\ 
NGC~4697 & 31.93 & NGC4731 & SBc & -20.98 &  8  \\
 & & NGC4775 & Scd & -20.60 &  \\ \\
NGC~5084 & 32.40 & NGC5084 & S0 & -21.33 & 12 \\ 
 & & NGC5087 & E-S0 & -20.65 &  \\ 
\hline \\
Fornax & 31.39$^a$ & NGC1399 & E & -21.04 & 161/147$^b$  \\ \\
Virgo & 30.92$^a$ & M87 & E & -21.48 & 76/53$^c$  \\ 
\hline
\end{tabular}
\ec
\begin{flushleft}
$^a$ Distance modulus derived from Cepheid distances (Freedman et al. 2001) \\
$^b$ Number of galaxies with positions within the 2$\degr$/1$\degr$.6 diameter field centered on NGC1399 
and velocities between 0 and 3000 km~s$^{-1}$ (the Fornax Cluster recession velocity limits) obtained from NED.\\  
$^c$ Number of galaxies with positions within the 2$\degr$/1$\degr$.6 diameter field centered on M87  
and velocities between -1000 and 3000 km~s$^{-1}$ (the Virgo Cluster recession velocity limits) obtained from NED.
\end{flushleft} 
\end{table*}

\section{Selection of Galaxy Groups}

We have chosen six galaxy groups at redshifts similar to the Fornax and Virgo Clusters. These redshifts 
(around 1500 km~s$^{-1}$) are sufficiently large to separate UCDs from Galactic stars in velocity, but not too high 
to require very long exposure times.
The group properties are summarized in Table~1.  
Column (2) is the distance modulus derived from the mean group radial velocity (as in Table 4 of   
Firth et al. 2006), corrected for large-scale motions using Equation A2 of Mould et al. (2000)\footnote{Note that the 
declinations of the Great Attractor and the Shapley Supercluster given 
in Table A1 of Mould et al. (2000) are negative, and the minus signs 
in their Equation A2 should all be positive.}.   
The Hubble constant is assumed to be $H_0 = 71$ km~s$^{-1}$ Mpc$^{-1}$. 
Columns (3) to (5) show the properties of the most luminous galaxies in the 2dF field observed.
Column (4) is the morphological type from HyperLeda\footnote{http://leda.univ-lyon1.fr} (Paturel et al. 2003). 
Column (5) is the absolute B-band magnitude, calculated using the total apparent corrected B-magnitude 
from HyperLeda and distance modulus from column (2). Column (6) is the number of galaxies (spectroscopically 
confirmed group members) with positions within 
the 2$\degr$ diameter field centered on the group centre of mass for Dorado, NGC1400/1407, NGC0681 
and 1$\degr$.6 diameter field for NGC4038, NGC4697, NGC5084.
The  numbers are based on Ferguson \& Sandage (1990) and Garcia (1993) galaxy group catalogues, supplemented with 
NED data and our 2dF galaxy observations (Firth et al. 2006).

We selected groups with a range of properties such as number of galaxies and a 
dominant central galaxy type. 
The groups have both early and late type central galaxies and may contain different numbers of UCDs if morphology 
is an indicator of the dynamical evolutionary state of the system.
The number of galaxies, spectroscopically confirmed group members,   
within the observed 2$\degr$/1$\degr$.6 diameter field centered on the group centre of mass  
ranges from 8 to 34 for individual groups.
If UCDs are stripped nuclei of early-type dwarf galaxies, the efficiency of UCD formation (the number of UCDs) 
should scale with the total mass of the group (Bekki et al. 2003) and therefore the number of galaxies. 
The same is true if UCDs are intra-cluster GCs: their number will correlate with the total mass of the group   
(Tonry \& Metzger 1997; McLaughlin 1999; Blakeslee, Bekki \& Yahagi 2006).
The number of UCDs in galaxy threshing hypothesis depends not only on the group mass, but also on the number of 
early-type dwarfs initially presented in the group (Bekki et al. 2003). 
Some of these early-type dwarfs will survive till nowadays.  
The NGC1400/1407 group contains many dE and dS0 dwarfs among spectroscopically confirmed members,  
all other groups do not (at least within the observed fields).
If UCDs are star clusters---GCs or YMGCs---formed in galaxies, 
the mass (luminosity) of the most massive (luminous) 
UCD should scale with the host galaxy mass (Kravtsov \& Gnedin 2005). 

One of our selected groups, NGC4038, 
includes the famous Antennae system (NGC4038/4039), an interacting pair, and 
is important in understanding UCDs as YMGCs in interacting 
systems. The group includes lots of galaxies with peculiar morphology: mergers, interacting 
systems and irregular type galaxies.   

The NGC5084 group is also of particular interest. NGC5084 is one of the most massive disk galaxies 
known with a mass of $M \sim 6 \times 10^{12} - 1 \times 10^{13} M_{\odot}$ (at a distance of 15.5 Mpc) and 
$M/L_B \gtrsim 200 \, M_{\odot}/L_{\odot}$ indicating a considerable amount of dark matter 
(Carignan et al. 1997). 
According to Carignan et al. (1997) this galaxy has survived the accretion of several 
satellites.

The NGC1400/1407 group is known by an extremely large difference in velocities between   
the second brightest member, NGC1400 (${\rm cz} \approx 600$ km~s$^{-1}$), and all other galaxies in the group, 
including the brightest member, NGC1407 (group mean velocity ${\rm cz} \approx 1700$ km~s$^{-1}$, 
Firth et al. 2006).
However, it was shown that NGC1400 is at the distance of the group (Gould 1993) and 
the difference in velocities was interpreted as the evidence for a large dark 
matter content (Quintana, Fouque \& Way 1994). Nevertheless, the spatial distribution and population size  
of the NGC1400 and NGC1407 globular cluster systems show no anomalies (Perrett et al. 1997).

\begin{table*}
\caption[2]{Observations.} 
\bc
\begin{tabular}{lcccccccc}
\hline
\\
Group & Date & \multispan2{\hfil 2dF field center\hfil} & $b_j$ & Total & Seeing & No. of objects & Completeness \\
 & & R.A.(J2000) & Dec.(J2000) &  & obs. time & & with measured & \\
 & & (h$\,$:$\,$m$\,$:$\,$s) & (d$\,$:$\,$m$\,$:$\,$s) & (mag) & (hr) & (arcsec) & velocities & (\%) \\
\hline
\\
Dorado & 2004 Nov 12--15 & 04:17:03.9 & -56:07:43 & 17.5 -- 20.5 & 8.75 & 1.4 -- 3.0 & 1196 & 43 \\
NGC~1400 & 2004 Nov 13--16 & 03:40:25.0 & -18:37:16 & 17.5 -- 20.5 & 8.50 & 1.2 -- 2.4 & $\,$~985 & 62 \\
NGC~0681 & 2004 Nov 12--15 & 01:49:49.7 & -10:03:05 & 17.5 -- 20.5 & 8.25 & 1.2 -- 3.0 & 1192 & 75 \\ 
NGC~4038 & 2005 Apr 02--05 & 11:59:57.2 & -19:16:21 & 17.5 -- 20.5 & 7.67 & 1.5 -- 2.0 & 1426 & 59 \\ 
NGC~4697 & 2005 Apr 02--05 & 12:53:55.1 & -06:15:41 & 17.0 -- 20.5 & 5.68 & 1.5 -- 2.0 & $\,$~816 & 46 \\
NGC~5084 & 2005 Apr 01--05 & 13:21:30.4 & -21:15:18 & 19.2 -- 21.0 & 8.96 & 1.5 -- 2.0 & $\,$~826 & 36 \\ 
\hline
\end{tabular}
\ec
\end{table*}

\begin{table*}
\caption[3]{New definite (in bold face) and possible members in the Dorado and NGC1400/1407 groups. 
cz is the heliocentric radial velocity. The best template Sp type is the spectral type of the template (star) 
which gave the highest R coefficient (given in the last column) when using the cross-correlation method 
(Tonry \& Davis 1979).} 
\bc
\begin{tabular}{cccccccc}
\hline
\\
Object & R.A.(J2000) & Dec.(J2000) & $b_j$ & $b_j - r$ & cz & Best template & R  \\
  & (h$\,$:$\,$m$\,$:$\,$s) & (d$\,$:$\,$m$\,$:$\,$s) & (mag) & (mag) & (km~s$^{-1}$) & Sp type & \\
\hline
\\
Dorado: &  &  &  &  &  &  &  \\
 1 & 04:16:24.55 & -56:37:19.7 & 19.7 & 1.1 & \,~585$\pm$71 & G6 V & 4.5 \\
 2 & 04:16:50.79 & -55:53:03.0 & 19.9 & 0.5 & \,~638$\pm$95 & F3 V & 4.6 \\
{\bf 3} & {\bf 04:13:16.91} & {\bf -55:46:26.0} & {\bf 20.2} & {\bf 0.9} & {\bf 1142$\pm$82} & {\bf G6 V} & {\bf 4.6}\\ \\
\,\,\,~~~NGC~1400: &  &  &  &  &  &  &  \\
 1 & 03:42:32.46 & -18:23:00.1 & 19.9 & 0.7 & \,~477$\pm$73 & F6 V & 4.3 \\
 2 & 03:41:54.81 & -18:08:37.8 & 19.5 & 0.9 & \,~652$\pm$65 & G6 V & 5.7 \\
\hline
\end{tabular}
\ec
\end{table*}

\section{Observations} 

Searches for UCDs were done with the Two Degree Field (2dF) multi-object spectrograph on 
the Anglo Australian Telescope in a single 2$\degr$ or 1$\degr$.6 
diameter field in each group centered on the group centre of mass. The coordinates of the group centers were 
taken from NED\footnote{The NASA/IPAC Extragalactic Database (NED) which is operated by the Jet Propulsion Laboratory, 
California Institute of Technology, under contract with the National Aeronautics and Space Administration.}. The source  
of these positions is Garcia (1993).
We had two observing runs: in November 2004 and in April 2005.
The summary of the observations is given in Table~2.

To improve our chances of finding UCDs, 
we defined a sample of objects looking similar to Fornax and Virgo UCDs: \\
(1) They are unresolved (star-like) in photographic plates. \\
(2) The magnitude limits for our targets were set so to match approximately 
the absolute magnitude range of the brightest Fornax and Virgo UCDs.   
We were unable to search for
  fainter UCDs because of the limited observing time allocated. \\ 
(3) A colour cut $b_j - r < 1.7$ was applied to remove Galactic M-dwarfs  
(no UCDs have been found redder than $b_j - r = 1.5$).

The UCD candidates for the November observing run---Dorado, NGC1400/1407 and NGC0681---were taken from 
the APM sky catalogues\footnote{http://www.ast.cam.ac.uk/$\sim$apmcat/} based on APM measurements of UK Schmidt 
telescope (UKST) blue and red photographic survey plates. We selected objects classified as ``stellar'' and ``merged''.
For the April observing run, UCD candidates were selected as star-like objects from the APM sky-survey catalogues 
for the NGC4038 and NGC5084 groups and SuperCOSMOS\footnote{http://www-wfau.roe.ac.uk/sss/index.html} scans of 
photographic UKST survey plates for the NGC4697 group. 
Targets were grouped by apparent magnitude, with exposures of about 45~min 
($b_j \lesssim 18.5$), 2~hr ($18.5 \lesssim b_j \lesssim 19.5$) and 
3~hr ($19.5 \lesssim b_j$) to obtain spectra with a similar signal-to-noise 
for all the targets. 
Poor weather conditions did not allow us to follow the observing plan for the NGC5084 group and to obtain data 
in the bright magnitude range ($b_j < 19.2$).
In the case of NGC4038, NGC4697 and NGC5084, we had a problem of too many targets per 2dF field due 
to the proximity of Galactic Plane. To observe a more complete sample 
within the limited observing time we restricted our observations to a 1$\degr$.6 diameter field for these groups.

The observing setup and data reduction procedure are 
identical to those used for the Fornax Cluster spectroscopic survey (Drinkwater et al. 2000b). 
We measured redshifts (radial velocities) for the UCD candidates by cross-correlation with a set of template spectra  
(see Drinkwater et al. 2000b for more details). Only velocities obtained with ${\rm R} \geq 3$ (Tonry \& Davis 1979) 
were accepted. 
The number of objects with measured velocities and completeness are given in Table~2.

We also obtained redshifts of galaxies, candidates for group members, simultaneously with the UCD candidate 
observations by using a small number of the available 2dF fibers to improve our knowledge  
about the groups themselves (Firth et al. 2006).

\section{Results}

Figure~1 represents the histogram of heliocentric radial velocities. 
The distribution of star-like objects from our observations is shown with the hatched histogram. 
The solid line histogram shows the distribution of galaxies, group members. The galaxy data were taken from 
Ferguson \& Sandage (1990) and Garcia (1993) galaxy group catalogues and supplemented with NED data and 
our 2dF galaxy observations (Firth et al. 2006).

There is no overlap between the star-like object and galaxy distributions for the NGC0681, NGC4038, NGC4697 and 
NGC5084 groups. No new group members (UCDs) were found in these groups. 
In the case of Dorado, one star-like object has a velocity of 1142 km~s$^{-1}$, which makes it a 
definite group member (UCD). Two star-like objects have velocities similar to the velocities of some galaxies in 
the group ($\sim 600 \, {\rm km~s}^{-1}$). However, these objects could be in the tail of stellar velocities 
once we allow for their largish errors. 
High-resolution imaging is needed to distinguish galaxies with ultra-compact 
morphology or globular clusters from Galactic stars. 
In the case of NGC1400/1407, two ``stars'' lie close to the NGC1400 galaxy in the velocity space   
($\sim 500 - 600 \, {\rm km~s}^{-1}$), but if we check their positions, neither of them seems associated with NGC1400. 
High-resolution imaging is required to make conclusions on the nature of these objects. 

The new definite and possible group members are listed in Table~3. Their positions in the galaxy groups are shown 
in Figure~2 with circles.  
The new definite member in Dorado is situated in intra-group space, far from any galaxies, and appears stellar in 
UKST photographic plates.
It has an absorption-line spectrum, similar to that of Fornax and Virgo UCDs and early-type dwarfs 
(Drinkwater et al. 2000a; Jones et al. 2006). 

As we did not observe all the candidate objects in each group, we have
estimated (95\% confidence) upper limits on the number of UCDs in each
group in the observed magnitude range (listed in Table~4). The upper limits
were estimated by using the binomial distribution to find the
(largest) UCD fraction among the candidates at which the probability
of finding no more UCDs than observed was 5\%.

\section{Discussion and Summary}

\begin{table}
\caption[4]{Upper limits on the number of UCDs (at 95\% confidence) in each
group in the observed magnitude range using the binomial distribution statistics.} 
\bc
\begin{tabular}{lc}
\hline
Dorado & 11.1/18.1$^a$ \\
NGC~1400 & $\,$~4.9/10.2$^b$ \\ 
NGC~0681 & $\,$~4.0  \\ 
NGC~4038 & $\,$~5.1  \\ 
NGC~4697 & $\,$~6.4  \\ 
NGC~5084 & $\,$~8.3 \\ 
\hline
\end{tabular}
\ec
\begin{flushleft}
$^a$ 1st number if we assume 1 UCD was detected; 2nd number if we assume 3 UCDs were detected \\
$^b$ 1st number if we assume no UCDs were detected; 2nd number if we assume 2 UCDs were detected
\end{flushleft} 
\end{table}

\begin{table}
\caption[5]{Predictions for the apparent magnitude of the most luminous UCD in each group.} 
\bc
\begin{tabular}{lc}
\hline
\\
Group & $b_j$ magnitude \\ 
\hline
\\
Dorado & 19.3 -- 19.6   \\
NGC~1400 & 18.0 -- 18.7  \\ 
NGC~0681 & 22.5 -- 22.6  \\ 
NGC~4038 & 21.8 -- 21.9  \\ 
 & 18.4 -- 18.6$^a$  \\
NGC~4697 & 22.1 -- 22.2  \\ 
NGC~5084 & 15.4 -- 16.7$^b$  \\ 
 & 19.7 -- 20.2$^c$  \\
\hline
\end{tabular}
\ec
\begin{flushleft}
$^a$  Here we assumed that UCDs are YMGCs such as found in Antennae system,  
and calculated the UCD absolute magnitude from its mass and mass-to-light ratio ($M/L$). $M/L$ was 
derived from SSP model predictions by Maraston (2005) for the YMGC age from Fall, Chandar \&  Whitmore (2005): 
$M/L \sim 0.03 \, M_{\odot}/L_{\odot,B}$ for the age $10^7$ yr.
The mass of the most massive UCD was estimated from the masses of NGC4038 and NGC4039 
and formula~8 of Kravtsov \& Gnedin (2005): $\sim 10^6 M_{\odot}$, which is consistent with the maximum mass  
of YMGCs in Antennae (e.g. Mengel et al. 2002). \\
$^b$ The estimation obtained from the mass of NGC5084 determined by Carignan et al. (1997). \\
$^c$ The estimation obtained from the virial mass of NGC5084, which was estimated from the internal velocity dispersion 
as given in HyperLeda and half-light radius as given in RC3.
\end{flushleft} 
\end{table}

Only one definite UCD candidate was detected in the six galaxy groups observed. 
To interpret this result we estimate the expected luminosity range of UCDs in groups 
in the context of the UCD formation scenarios.
In particular, we consider the hypothesis that UCDs are globular clusters formed in galaxies. 
This is the only hypothesis for which we have {\it quantitative} theoretical predictions.

If UCDs are GCs formed in galaxies (and    
subsequently escaped their host galaxy potential as e.g. suggested by Bekki \& Yahagi 2006),
then we can use predictions by Kravtsov \& Gnedin (2005). They found that  
the most massive cluster contributes a significant fraction of the total cluster mass and that 
the mass of the most massive cluster correlates with the mass of its host galaxy.
This picture is consistent with what we found in both Fornax and Virgo galaxy clusters (Drinkwater et al. 2004;  
Evstigneeva et al. 2007; Hilker et al. 2007; Gregg et al. 2007): 
there is one very massive (luminous) UCD and there are other, less massive (less luminous) ones. 
Also, the fainter in luminosity we go, the more UCDs we have. 
Therefore, if we find the luminosity (mass) of the most luminous (massive) UCD for each group, we immediately 
find the luminosity (mass) of all other UCDs in the group: they will be fainter (less massive) than the most 
luminous UCD.  

Formula 8 of Kravtsov \& Gnedin (2005) gives    
the relation between the mass of the most massive GC and mass of its host galaxy. 
The formula works very well for Fornax UCDs \& the NGC1399 galaxy and Virgo UCDs \& 
the M87 galaxy (Evstigneeva et al. 2007). We used this relation to predict the mass  
of the most massive UCD for each group. Most massive UCDs are obviously formed 
in most massive galaxies (according to the adopted hypothesis).
The mass values for the most massive galaxies in the groups were taken from: 
NGC1407 and NGC1400 -- Quintana et al. (1994); NGC5084 -- Carignan et al. (1997); 
NGC4038 and NGC4039 -- Amram et al. (1992); NGC4027 -- Phookun et al. (1992);  
NGC0681 -- Kyazumov \& Barabanov (1980); NGC4731 -- Gottesman et al. (1984);  
NGC1553 and NGC1549 -- virial masses, estimated from the internal velocity dispersion 
as given in HyperLeda and half-light radius as given in RC3 (de Vaucouleurs et al. 1991).
The most massive galaxies in the groups are usually the most luminous ones, but there are exceptions.
In the NGC4038 group, for example, NGC4027 seems to be the most massive galaxy.  
To convert the UCD masses into luminosities, the mass-to-light ratio $M/L_B=3$ was taken 
(to reproduce the mass range of Fornax and Virgo UCD from 
Hilker et al. (2007) and Evstigneeva et al. (2007) for their $b_j$ luminosities). The luminosities 
were in turn converted into apparent $b_j$ magnitudes using the group distance moduli from Table~1. 
Table~5 gives the estimated $b_j$ magnitude for the most luminous UCD in each group.
The uncertainty is due to the scatter around the relation described by formula 8 of Kravtsov \& Gnedin (2005).

We can now compare our observational results---1 definite plus 2
possible UCD candidates in Dorado and 2 possible UCD candidates in the
NGC1400 group---with the predictions given in Table~5.  
According to the predictions, we expect to find UCDs in the Dorado, NGC1400 and NGC5084 groups. 
This broadly agrees with the observational results (the exception is NGC5084). 
We would also expect to find UCDs in the NGC4038 group if they were the likes of YMGCs in Antennae 
(see notes for Table~5), but we did not. 
It is possible that we did not find UCDs in NGC5084, because the dominant galaxy (NGC5084) was near  
the edge of the observing field and our coverage was incomplete. 
The same can be said about YMGCs in the NGC4038 group and the Antennae system. 

To make better tests of this and other hypotheses for UCD formation in the galaxy group
environment, we clearly need deeper observations (compare the
predictions in Table~5 with our observational limits). This would
require much larger allocations of observing time, although the
efficiency of the UCD candidate selection could be substantially
improved by using multicolour CCD photometry for the input catalogues
(e.g. Firth et al. 2007, in preparation). \\

In this paper we have presented the results of a search for UCDs 
in six different galaxy groups. We found 1 definite plus
2 possible UCD candidates in the Dorado group and 2 possible UCD candidates in
the NGC1400 group. No UCDs were found in the other groups. We compared
these results with predicted luminosities of UCDs in the
groups according to the hypothesis that UCDs are globular clusters formed in galaxies. 
The predictions broadly agree with the observational results.

\begin{figure*}
\centerline{\psfig{file=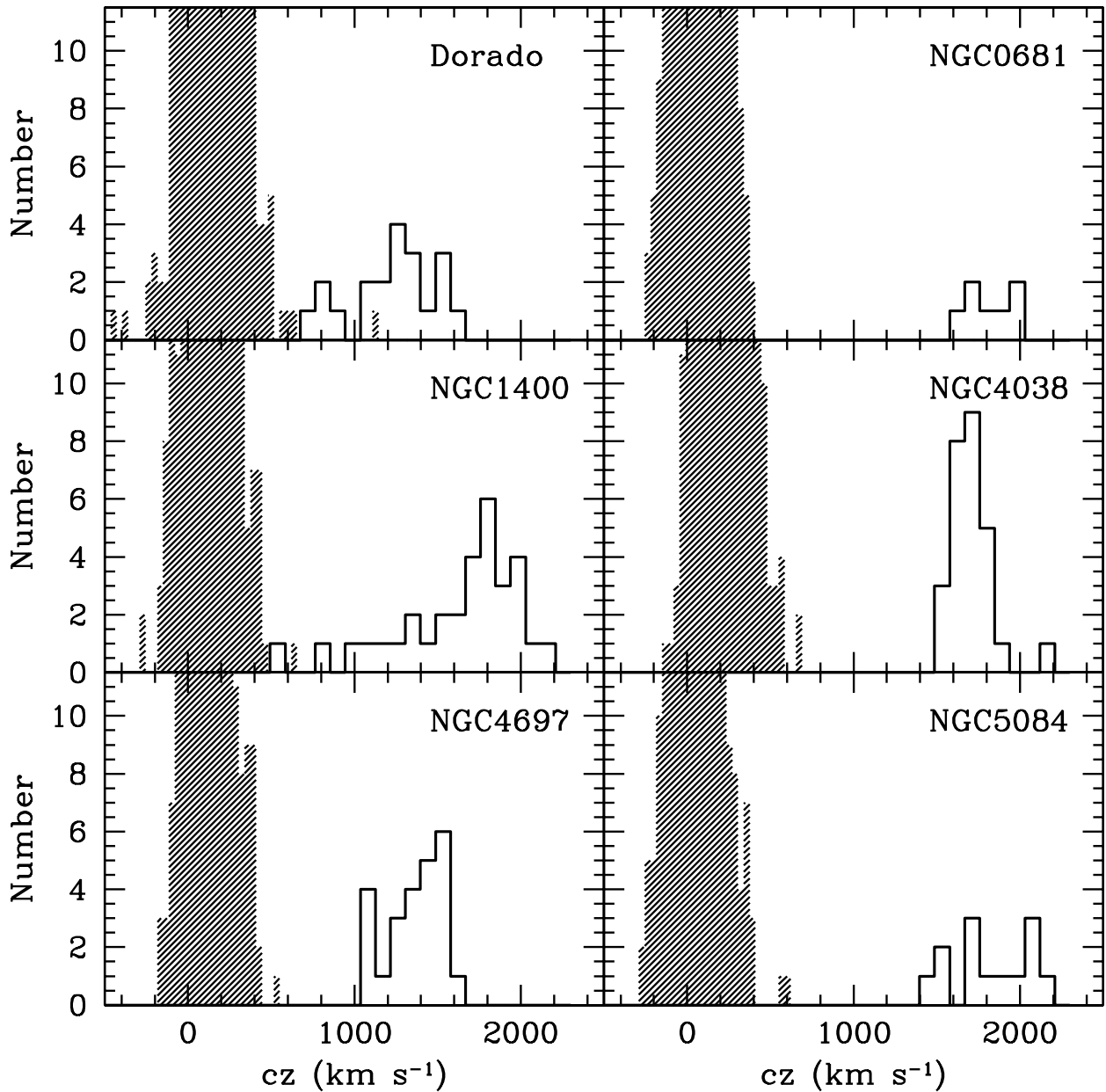,width=18cm,clip=}}
\caption{Histograms of radial velocities. The distribution of star-like objects from our observations 
is shown with hatched histogram. Solid line shows the distribution of galaxies, group members, from 
Ferguson \& Sandage (1990) and Garcia (1993) galaxy group catalogues, supplemented with NED data and 
our 2dF galaxy observations (Firth et al. 2006).}
\end{figure*}  

\begin{figure*}
\centerline{\psfig{file=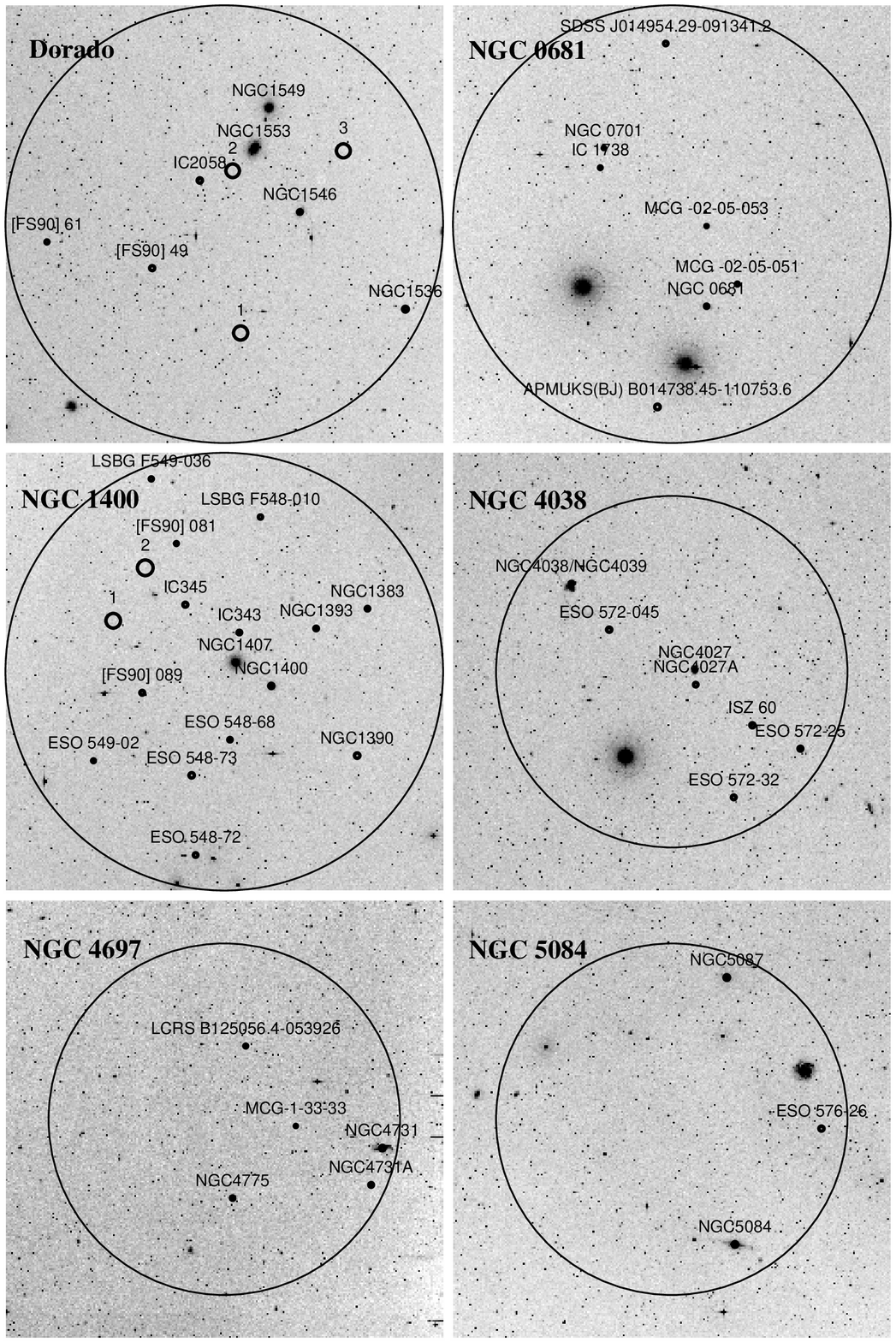,width=15.0cm,clip=}}
\caption{DSS images of six galaxy groups with some galaxies (spectroscopically confirmed group members) labeled.
All the images have a size of $2\degr \times 2\degr$ and were retrieved from the Canadian Astronomy Data 
Centre (http://cadcwww.dao.nrc.ca/). 
The circle represents the 2$\degr$/1$\degr$.6 diameter field centered on the group centre of mass.
Definite and possible UCDs in Dorado and NGC1400/1407 are labeled as in Table~3.} 
\end{figure*}

\section*{Acknowledgements}

EAE and MJD acknowledge support from the Australian Research Council.
We thank Rob Sharp for his assistance during observations.
We are grateful to the referee for helpful suggestions, which have improved this paper.

\bsp

\end{document}